\documentclass[12pt,letterpaper]{article} 
\pdfoutput=1
\usepackage[includeheadfoot,
            marginratio={1:1,2:3}, 
            width=412pt, 
            height=688pt,]{geometry}

\usepackage{amsmath}
\usepackage{amsfonts}
\usepackage{amssymb}
\usepackage{stmaryrd}
\usepackage{graphicx}
\usepackage{cite}
\usepackage{hyperref}

\newcommand{\nc}{\newcommand}
\nc{\lb}{\llbracket}
\nc{\rb}{\rrbracket}
\nc{\gl}{\llbracket}
\nc{\gr}{\rrbracket}

\newcommand{\eq}[1]{\begin{equation}
                     \begin{split} #1 \end{split}
                     \end{equation}}

\newcommand{\ov}{\overline}

\newcommand{\op}{\hspace{1pt}}
\newcommand{\parag}{\,\begin{matrix} \gtrsim \\[-0.3cm]{}_p \end{matrix}\,}
\def\mff{\mathfrak f}

\allowdisplaybreaks[2]
\numberwithin{equation}{section}

\begin{document}

\vspace*{-1.5cm}
\begin{flushright}
  {\small
  MPP-2015-249\\
  }
\end{flushright}

\vspace{1.5cm}
\begin{center}
 {\LARGE The Flux-Scaling Scenario: \\ [0.2cm]
De Sitter Uplift and Axion Inflation }
\vspace{0.4cm}

\end{center}

\vspace{0.35cm}
\begin{center}
  Ralph Blumenhagen$^{1}$, Cesar Damian$^{1}$,  Anamar\'{\i}a Font$^{2}$,\\
  Daniela Herschmann$^{1}$, Rui Sun$^{1}$
\end{center}

\vspace{0.1cm}
\begin{center} 
\emph{$^{1}$ Max-Planck-Institut f\"ur Physik (Werner-Heisenberg-Institut), \\ 
   F\"ohringer Ring 6,  80805 M\"unchen, Germany } \\[0.1cm] 
\vspace{0.25cm} 
\emph{$^{2}$ 
Departamento de F\'{\i}sica, Centro de F\'{\i}sica Te\'orica y Computacional \\
Facultad de Ciencias, Universidad Central de Venezuela\\
 A.P. 20513, Caracas 1020-A, Venezuela}\\
\vspace{0.25cm} 

\vspace{0.2cm}

 \vspace{0.5cm} 
\end{center} 

\vspace{1cm}


\begin{abstract}
Non-geometric flux-scaling vacua provide promising starting points to realize axion monodromy inflation 
via the F-term scalar potential. 
We show that these vacua can be uplifted to Minkowski and de Sitter by adding an $\overline{{\rm D}3}$-brane or 
a D-term containing geometric and non-geometric fluxes. 
These uplifted non-supersymmetric models are analyzed with respect to
their potential to
realize axion monodromy inflation self-consistently. Admitting rational
values of the fluxes, we construct examples with
the required hierarchy of mass scales.
\end{abstract}

\clearpage




\section{Introduction}
\label{sec:intro}

Motivated by realizing single field F-term axion monodromy inflation \cite{Marchesano:2014mla,Hebecker:2014eua,Blumenhagen:2014gta},
while taking closed string moduli stabilization into account,
a scheme of high scale supersymmetry breaking was proposed in
\cite{Blumenhagen:2015kja}.
The inflaton was an axion receiving a (flattened) polynomial potential from 
a tree-level background flux, thus achieving large field inflation
with an  observable tensor-to-scalar ratio and an inflationary scale
of the order of the GUT scale and with an inflaton mass of order
$10^{13}\, {\rm GeV}$. 
It is worth to 
emphasize that, after its inception in \cite{Kaloper:2008fb,Kaloper:2011jz,Silverstein:2008sg,McAllister:2008hb}, 
the stringy realization of axion monodromy inflation has
become an active area of research (see e.g. \cite{Baumann:2014nda,Westphal:2014ana} for 
reviews). Just to mention a few developments, in \cite{Arends:2014qca} the axion responsible for inflation was identified
with a deformation
modulus of a ${\rm D}7$-brane, whereas in \cite{McAllister:2014mpa,Franco:2014hsa} the axion was related to the 
$B$-field from the NS-NS sector
integrated over a non-contractible internal two cycle. 
In \cite{Hassler:2014mla} non-geometric fluxes were included in the effective theory identifying 
the K\"ahler modulus with the inflaton. 
Other scenarios realize axion inflation in warped resolved
conifolds \cite{Kenton:2014gma}, which suffers from a too small string scale for a large axion decay constant \cite{Kooner:2015rza}. 
The case of chaotic inflation with axionic-like fields considering the
backreaction of the heaviest moduli has been worked out in \cite{Buchmuller:2015oma}. Another attempt to embed chaotic inflation
is  \cite{Ibanez:2014kia} 
where the axion was identified with either a Wilson line or the position modulus of a ${\rm D}$-brane containing the MSSM.
In the
framework of F-theory \cite{Grimm:2014vva}, an axion-like field serves as inflaton for natural inflation. 
Special points in the moduli space for which the complex structure moduli can drive axion monodromy inflation 
were investigated in \cite{Garcia-Etxebarria:2014wla}.

Since for single field
inflation,  the inflaton should be the lightest scalar field, all the other
moduli should better acquire their masses already at tree-level.
For type IIB orientifold compactifications on Calabi-Yau (CY) three-folds
this means in particular that all closed string moduli, namely
the axio-dilaton as well as the complex structure and K\"ahler moduli,
should be stabilized by geometric and non-geometric fluxes.
Closed string moduli stabilization with solely
fluxes was discussed in \cite{Blumenhagen:2015kja} while  its application
to axion inflation was further elucidated in
\cite{Blumenhagen:2015qda}.

One of the main results of \cite{Blumenhagen:2015kja} is that by turning on $n+1$ 
fluxes for $n$
moduli, the resulting F-term scalar potential admits so-called scaling
type non-supersymmetric  AdS minima with the desired  properties.
Here scaling type means that the values of the moduli in the minimum,
as well as all the mass scales, are determined by ratios of products of 
fluxes, thus allowing for parametric control of these quantities.
This is important in order to argue for the self-consistency
of the moduli stabilization scheme, i.e. that eventually the moduli
are stabilized  in their perturbative regime and that, 
e.g. the moduli masses are separated from the string and Kaluza-Klein
scales.

Conceptually, the induced F-term scalar potential is related to the one of $N=2$ gauged
supergravity by an orientifold projection breaking $N=2$ down to $N=1$ \cite{D'Auria:2007ay}.
Recently, it was explicitly shown in \cite{Blumenhagen:2015lta} that the same 
potential also arises by appropriate dimensional reduction of double field
theory on a CY three-fold equipped with fluxes. In fact, it turns out that
the latter also includes a D-term potential that emerges when
there are abelian gauge fields present coming from the dimensional reduction of
the R-R four-form on an orientifold even  three-cycle of the CY \cite{Robbins:2007yv}.

It is important to note that, throughout the work
\cite{Blumenhagen:2015kja}, 
it was assumed that the flux-scaling AdS vacua
could be uplifted to Minkowski or to de Sitter vacua,
for instance by introducing an $\ov{{\rm D}3}$-brane as in the KKLT scenario \cite{Kachru:2003aw}.
As a fairly new and significant development, it has been recently pointed out that
this often employed $\ov{{\rm D}3}$-brane uplift mechanism can be described
within supergravity by a nilpotent superfield \cite{Kallosh:2014wsa,Bergshoeff:2015jxa,Kallosh:2015nia} and
the vacua are argued to be metastable\cite{Polchinski:2015bea}.
However, in \cite{Blumenhagen:2015kja},
for one concrete example it was shown that a naive uplift of flux-scaling AdS vacua by
introducing an $\ov{{\rm D}3}$-brane in a warped throat does not work.
Indeed, by increasing the warp factor, the
minimum got destabilized before the cosmological constant
vanished. However, for string theory to provide a reliable description of inflation, 
it has to explain the cosmological constant in a self-consistent compactification.

In the past years,  potential realizations of dS vacua in
string theory  have been intensively studied from different
 perspectives
\cite{Kachru:2003aw,Achucarro:2006zf,Gray:2008zs,Krippendorf:2009zza,Danielsson:2012by,Blaback:2013ht,
Damian:2013dq,Damian:2013dwa,Kallosh:2014oja,Rummel:2014raa,Blaback:2015zra}.
Both analytical and numerical  approaches have been
followed to construct metastable dS vacua. 
Moreover, as a useful guide,  no-go theorems have been derived
in the context of the type II 
\cite{Hertzberg:2007wc,Haque:2008jz,Flauger:2008ad,Danielsson:2009ff,
deCarlos:2009fq,Caviezel:2009tu,Wrase:2010ew,Shiu:2011zt,Borghese:2012yu}
and heterotic \cite{Green:2011cn,Gautason:2012tb,Kutasov:2015eba}
superstrings.

One of the loopholes of these no-go theorems is the restriction
of the fluxes to those  visible in supergravity. However, by
arguments based on T-duality \cite{Shelton:2005cf,Shelton:2006fd} and the 
developments in generalized geometry and double field theory \cite{Aldazabal:2013sca,
Berman:2013eva,Andriot:2011uh,Geissbuhler:2013uka,Blumenhagen:2013hva}
it has  become clear that there might also exist so-called non-geometric fluxes. 
For instance,   the  $STU$-models
\cite{Font:2008vd,Guarino:2008ik,deCarlos:2009qm,Aldazabal:2011yz,Dibitetto:2011gm}
were analyzed in much detail for realizations of   dS vacua via
the introduction of T- and S-dual non-geometric fluxes.

Since the question of uplifting is clearly a very important unsettled issue in 
the flux-scaling scenario,
it is the purpose of this paper to investigate this problem more closely.
First, for the $\ov{{\rm D}3}$-brane
case we will find that adding the tension of this brane to the
flux induced F-term  potential can  lead to new flux-scaling solutions
that are of Minkowski/de Sitter type.  
Second, as mentioned above, for $h^{21}_+>0$ there is an additional
positive semi-definite D-term contribution to the scalar 
potential \cite{Robbins:2007yv, Blumenhagen:2015lta}
that in principle could also help with increasing the cosmological
constant at the minimum. We will show that this alternative 
also works.
Let us emphasize that 
these are not continuous uplifts of
initial  AdS minima, but just new minima lying on a different branch
in the landscape.

As mentioned,  the motivation for moduli stabilization in the flux-scaling scheme
was the stringy realization of axion monodromy inflation. Therefore,
having now two possible ways of uplifting available, we also
revisit the problem of realizing axion monodromy inflation.
We still find that for integer
quantized fluxes, it is persistently
difficult  to obtain  all mass scales in the right order, namely
\eq{
\label{hierarchy}
M_{\rm s} > M_{\rm KK} > M_{\rm inf}> M_{\rm mod} > H_{\rm inf} >
M_{\rm \theta} \, , \nonumber
}
where $\theta$ denotes the inflaton.
However, it is known that the perturbative corrections to the
prepotential of the complex structure moduli lead to a redefinition
of the fluxes  so that some of them become rational numbers.
Phenomenologically scanning over such  rational   values,
we identify  a model in which the above hierarchy is indeed fulfilled.

This paper is organized as follows:
In section \ref{sec:fluxscaling} we briefly review type IIB 
orientifolds on Calabi-Yau three-folds with various geometric 
and non-geometric fluxes turned on. 
In the main section \ref{sec:uplift} we present examples of uplifted 
flux-scaling vacua. We discuss one model with an $\overline{{\rm D}3}$-brane uplift 
and another with a D-term uplift. We also show
that by changing the warp factor for the former example, one can 
interpolate between  AdS and  dS  vacua.
In section \ref{sec:ami} we analyze the realization of  axion
monodromy inflation in the model with D-term-uplift.

\section{The flux-scaling scenario}
\label{sec:fluxscaling}

In this section, we first review the salient features of the moduli
stabilization scheme introduced in \cite{Blumenhagen:2015kja}. 
 For more details of this
construction we refer the reader to the original literature.

The starting point are 
orientifolds of the type IIB superstring compactified on Calabi-Yau 
three-folds with non-vanishing (non-)geometric fluxes turned on.
Such models have indeed been investigated before \cite{Grana:2005ny,Benmachiche:2006df,Grana:2006hr,Micu:2007rd,Palti:2007pm}.
The orientifold projection is $\Omega_{\rm P} (-1)^{F_L} \sigma$ where
$\sigma$ acts such that there are ${\rm O}7$- and ${\rm O}3$-planes.
For vanishing fluxes, the massless spectrum comprises 
$h^{1,1}_+$ complexified K\"ahler moduli $T_\alpha$, $h^{1,1}_-$
purely axionic  moduli $G^a$, $h^{2,1}_-$ complex structure moduli
$U^i$ and $h^{2,1}_+$   abelian gauge fields $A_j$ resulting from the
dimensional reduction of the R-R four-form $C_4$ on three-cycles
of the CY \cite{Grimm:2004uq}. In addition the dilaton and the R-R 0-form
give the chiral axio-dilaton, defined as $S=e^{-\phi} - i C_0$ in our conventions.

The various fluxes appear  in a twisted differential acting on $p$-forms. This differential contains
the constant fluxes $H$, $F$, $Q$ and $R$, and is given by
\eq{
  \label{d_operator_01}
  \mathcal D = d - H\wedge\: - F\circ\: - Q\bullet\: - R\,\llcorner \,,
}
where the operators entering in \eqref{d_operator_01} act as 
\eq{
  \renewcommand{\arraystretch}{1.2}
  \arraycolsep3pt
  \begin{array}{l@{\hspace{7pt}}c@{\hspace{12pt}}lcl}
  H\,\wedge & :& \mbox{$p$-form} &\to& \mbox{$(p+3)$-form} \,, \\
  F\,\circ & :& \mbox{$p$-form} &\to& \mbox{$(p+1)$-form} \,, \\
  Q\,\bullet & :& \mbox{$p$-form} &\to& \mbox{$(p-1)$-form} \,, \\  
  R\,\llcorner & :& \mbox{$p$-form} &\to& \mbox{$(p-3)$-form} \,.  
  \end{array}
}
For the different forms in a CY three-fold this action can be specified by \cite{Grana:2006hr}
\eq{\label{deffluxes}
\arraycolsep2pt
\begin{array}{lcl@{\hspace{1.5pt}}l@{\hspace{40pt}}lcr@{\hspace{1.5pt}}l}
\mathcal D\alpha_\Lambda &=& q_{\Lambda}{}^{ A} \omega_{ A}
&+\,  f_{\Lambda \op A}\tilde\omega^{ A}\,,
&
\mathcal D\beta^\Lambda &=& \tilde q^{\Lambda \op A} \omega_{ A}
&+ \,\tilde f^{\Lambda}{}_ { A} \tilde\omega^{ A}\,, 
\\[8pt]
\mathcal D\omega_{ A}&=& -\tilde f^{\Lambda}{}_{ A} \alpha_\Lambda &+  \,
f_{\Lambda  A}\beta^\Lambda\,,
&
\mathcal D\tilde\omega^{ A} &=& \tilde q^{\Lambda\op  A} \alpha_\Lambda &-
\,q_{\Lambda}{}^{ A} \beta^\Lambda\,.
\end{array}
}
with $\Lambda=0,\ldots,h^{2,1}$ and $A=0,\ldots, h^{1,1}$. 
For the $H$- and $R$-flux we further use the conventions  
\eq{
\label{fluxzerocomp}
\arraycolsep2pt
\begin{array}{lcl@{\hspace{70pt}}lcl}
f_{\Lambda \op0}&=&r_\Lambda\,, & \tilde f^{\Lambda}{}_0&=&\tilde r^\Lambda\,,\\[5pt]
q_{\Lambda}{}^0&=&h_\Lambda\,,& \tilde q^{\Lambda\op 0}&=&\tilde h^\Lambda\, .
\end{array}
}
We also define $\tilde \omega^0=1$, and $\omega_0= \sqrt g \op d^6x/{\mathcal V}_{\mathcal M}$, where
${\mathcal V}_{\mathcal M}=\int_{\mathcal M} \sqrt g \op d^6x $ is the volume of the CY three-fold ${\mathcal M}$.

Imposing the nilpotency condition of the form $\mathcal D^2=0$ leads to  
Bianchi identities for the fluxes. In this way we obtain 
\eq{
\label{Bianchi}
\arraycolsep2pt
\begin{array}{lcl@{\hspace{50pt}}lcl}
0&=&\tilde q^{\Lambda \op A} \tilde f^\Sigma{}_{ A} -  \tilde f^\Lambda{}_{ A} \tilde q^{\Sigma \op  A}\,,
&
0&=& q_{\Lambda}{}^{ A}  f_{\Sigma\op  A} -  f_{\Lambda \op  A}  q_{\Sigma}{}^ { A}\,,
\\[7pt]
0&=&q_{\Lambda}{}^{ A}  \tilde f^\Sigma{}_{ A}  -   f_{\Lambda\op  A}  \tilde q^{\Sigma \op A}\,,
&
0&=& \tilde f^\Lambda{}_{ A} q_{\Lambda}{}^{ B} - f_{\Lambda\op A} \tilde q^{\Lambda\op   B}\,.
\\[7pt]
0&=&\tilde f^\Lambda{}_{ A} f_{\Lambda\op  B}- f_{\Lambda \op A} \tilde f^\Lambda{}_{ B}\,,
&
0&=& \tilde q^{\Lambda \op A} q_{\Lambda}{}^{ B}- q_{\Lambda}{}^{ A}  \tilde q^{\Lambda\op  B}\,.
\end{array}
}
Implementing the orientifold projection, the invariant fluxes are
\eq{
\label{op_02}
\renewcommand{\arraystretch}{1.2}
\begin{array}{l@{\hspace{6pt}}c@{\hspace{18pt}}llll}
{\mathfrak F}&:& && {\mathfrak f}_{\lambda}\,, & \tilde {\mathfrak f}^{\lambda} \,,\\
H&:& && h_{\lambda}\,, &  \tilde h^{\lambda} \,,\\
F&:& f_{\hat \lambda \,\alpha}\,, &  \tilde f^{\hat\lambda}{}_{\alpha}\,, & f_{\lambda \,a}\,, &  \tilde f^{\lambda}{}_{a}\,, \\
Q&:& q_{\hat\lambda}{}^a\,, & \tilde q^{\hat\lambda\,a}\,, & q_{\lambda}{}^{\alpha} \,, & \tilde q^{\lambda\,\alpha}\,, \\
R&:& r_{\hat\lambda} \,, & \tilde r^{\hat\lambda} \,.
\end{array}
}
where $\lambda=0,\ldots,h^{21}_-$, 
$\hat\lambda=1,\ldots,h^{21}_+$, \mbox{$\alpha=1,  \ldots, h_+^{1,1}$} and  \mbox{$a=1,  \ldots, h_-^{1,1}$}. 
Note that in \cite{Blumenhagen:2015kja}, the construction was restricted  to the case
$h^{21}_+=0$, whereas here we also consider $h^{21}_+>0$. 
In fact, as shown in \cite{Blumenhagen:2015lta}, the fluxes with index $\lambda$
contribute to an F-term scalar potential whereas the fluxes with
index $\hat\lambda$ contribute to a positive definite D-term
potential.

For moduli stabilization, we allow 
all orientifold even fluxes, only subject to the Bianchi identities.
The superpotential generating the F-term potential takes the form \cite{Benmachiche:2006df,Grana:2006hr}
\eq{
  \label{s_pot_02}
  W= \int_{\mathcal M} \Bigl[ \,\mathfrak F + \mathcal D \op\Phi^{\rm ev}_{\rm c} \Bigr]_{3} \wedge \Omega\,
}
with the complex multiform 
$\Phi^{\rm ev}_{\rm c} = i\op S -i\op G^a \omega_a -i \op T_{\alpha} \op\tilde\omega^{\alpha}$.
Using \eqref{deffluxes} the superpotential  can be further evaluated as
\eq{
\label{thebigW}
   W=&
  -\bigl({\mathfrak f}_\lambda  X^\lambda -\tilde {\mathfrak f}^\lambda  F_\lambda  \bigr) 
  +i\op S \big( h_\lambda  X^\lambda - \tilde h^\lambda  F_\lambda \bigr) \\[4pt]
  &+i\op G^a  \bigl( f_\lambda{}_a   X^\lambda - \tilde f^\lambda{}_a  F_\lambda \bigr) 
  -i\op T_{\alpha}  \bigl( q_\lambda{}^\alpha   X^\lambda - \tilde q^\lambda{}^\alpha  F_\lambda\bigr)\,.
}
where the periods $X^\lambda, F_{\lambda}$  of the holomorphic 3-form $\Omega$ are computed from the tree-level
cubic prepotential ${F}={1\over 6} d_{ijk} X^i X^j X^k/X^0$ of the CY
three-fold\footnote{The generically present subleading polynomial corrections to this cubic form will be
  considered later.}.  Specifically, $\Omega$ has the expansion
\mbox{$\Omega =X^\lambda \alpha_\lambda - F_\lambda \beta^\lambda$}.

The tree-level K\"ahler potential in the large complex structure limit
can be expressed as \cite{Grimm:2004uq}
\eq{
  \label{k_pot}
      K=-\log\left(-i\int_{\mathcal M} \Omega\wedge \ov{\Omega}\right)-\log\bigl(S+\ov S\bigr)
     -2\log {\cal V} \,.
}
Here ${\mathcal V} = \frac16 \kappa_{\alpha \beta\gamma} t^\alpha t^\beta t^\gamma$
denotes the volume of the CY three-fold in Einstein frame. For future reference we also record
the expansions of the K\"ahler and NS-NS 2-forms, respectively \mbox{$J=e^{\phi/2} t^\alpha \omega_\alpha$} and 
\mbox{$B_2 = b^a \omega_a$}.

In \cite{Blumenhagen:2015lta}, it was explicitly shown that the F-term
scalar potential 
\eq{
  \label{f_pot}
  V_F = \frac{M_\text{Pl}^4}{4 \pi} \,\, e^{K} \Bigl( K^{I\ov J} D_IW D_{\ov J}\ov W - 3 \op\bigl|W\bigr|^2 \Bigr) \,,
}  
resulting from the K\"ahler potential and superpotential reviewed above,
is related to the one obtained via
dimensional reduction of double field theory on a Calabi-Yau three-fold
with (non-)geometric fluxes. 
Moreover, the potential is related to  $N=2$ gauged
supergravity \cite{D'Auria:2007ay}.
More concretely, taking the orientifold projection the latter scalar
potential splits into three pieces
\eq{
\label{potentialfull}
         V=V_F+V_D+V^{\rm NS}_{\rm tad}
}
where $V_F$ is precisely the F-term scalar potential \eqref{f_pot}. $V^{\rm NS}_{\rm tad}$
is the NS-NS tadpole that will be cancelled against the tension of the
branes and orientifold planes, once R-R tadpole cancellation is taken
into account. $V_D$ is an additional D-term potential 
\eq{
  \label{dtermpot}
   V_D &= -\frac{M^4_{\rm Pl}}{2} \, \Bigl[ ({\rm Im}\, \mathcal N)^{-1} \Bigr]^{\hat\lambda\hat\sigma} \op
   D_{\hat\lambda} \op D_{\hat\sigma} \\[4pt]
}
that results from the abelian
gauge fields for $h^{2,1}_+>0$. 
Adjusting the results in \cite{Blumenhagen:2015lta} to the present conventions,
the D-terms $D_{\hat\lambda}$ in Einstein frame are  given by
\eq{
\label{res_054}
D_{\hat{\lambda}} =
 \frac{1}{{\mathcal V}}\left[ 
  -r_{\hat\lambda}\, \big( e^\phi {\cal V}-\tfrac{1}{2}\op \kappa_{ \alpha a b} \op
    t^\alpha b^a b^b\Big) - q_{\hat\lambda}{}^{a}\, \kappa_{a \alpha  b}\op
     t^\alpha b^b + f_{{\hat\lambda} \alpha}\, t^\alpha \right].
}
We have set
$\tilde r^{{\hat\lambda}}=\tilde q^{{\hat\lambda} a}=\tilde f^{{\hat\lambda}}{}_{\alpha}=0$.

In \cite{Blumenhagen:2015kja}, assuming $h^{21}_+=0$, the F-term  scalar potential $V_F$ was investigated in
detail, and particular attention was paid to so-called scaling
type minima,  in which $W$ contained only $n+1$ terms for a model
with  $n$  superfields. This ansatz led to solutions where the fixed
moduli, as well as the resulting moduli mass scales, could be expressed
as simple quotients of fluxes. This allowed to gain parametric
control over certain mass scales which was important for the
realization of F-term axion monodromy inflation. All scaling 
vacua of this type were stable non-supersymmetric AdS minima, for which
the existence of an uplift to Minkowski/de Sitter was just assumed.
However,  for a simple concrete model it was shown that a simple
uplift \`a la KKLT does not really work, as the additional $\ov{{\rm D}3}$-brane
contribution to the scalar potential destabilized the vacuum.
In the following section, we will show that for  concrete simple examples
Minkowski/de Sitter minima exist featuring  also the nice scaling type
behavior. 

\subsubsection*{Non-geometric S-dual $P$-form fluxes}

After adding the non-geometric $Q$-fluxes, the superpotential (\ref{thebigW}) is no longer covariant under S-duality
transformations. It has been proposed that this covariance can be restored by including non-geometric $P$-fluxes, which
transform together with the $Q$-fluxes as a doublet of the
$SL(2,\mathbb{Z})$ duality group \cite{Aldazabal:2006up}. 
Similar to the $Q$-flux, the $P$-flux is defined as a map
\eq{P \bullet \; : \; p{\rm -form} \rightarrow (p-1){\rm -form} \, ,
}
and the action of $P$ on the symplectic basis is specified by
\eq{
\label{pflux1}
\arraycolsep2pt
\begin{array}{lcl@{\hspace{70pt}}lcl}
- P\bullet \alpha_{\Lambda} 	&=& 
p_{\Lambda}^{A}\,, 	& - P \bullet \beta^{\Lambda} 	&=& \tilde{p}^{\Lambda A}\omega_{A}\,,\\[5pt]
- P\bullet \omega_{A} 		
&=& 0\,,		& - P \bullet \tilde{\omega}^{A}&=& -p^{\Lambda A}\alpha_{\Lambda} + p_{\Lambda}^A \beta^{\Lambda}\, .\\
\end{array}
}
The extended superpotential is derived requiring that it transforms properly under S-duality.
Taking also into account the geometric moduli $G^a$ it is given by  \cite{Blumenhagen:2015kja}
\eq{
W' =\int_{\mathcal{M}} \big[ \mathfrak F + \mathcal D \Phi_{\rm c}^{\rm ev} + 
T_{\alpha} S \left( P\bullet \tilde{\omega}^{\alpha}\right)
+ \frac{1}{2} \kappa_{\alpha b c}G^b G^c \left( P\bullet \tilde{\omega}^{\alpha} \right) \big]_3 \wedge \Omega_3 \, ,
}
which after  integrations yields
\eq{
W' = W + \left(S T_{\alpha} +\frac{1}{2}\kappa_{\alpha b c}G^b G^c \right) 
\left( p_\lambda^{\alpha} X^\lambda - \tilde{p}^{\lambda \alpha}F_{\lambda} \right) \, ,
}
where $W$ is shown  in \eqref{thebigW}.

In this paper we will restrict attention to  examples with $h^{1,1}_- = 0$ so that the geometric 
$G^a$ moduli contribution to the scalar potential
is absent. The Bianchi identities in this case were discussed in \cite{Aldazabal:2006up}. For our purposes
we can take a pragmatic approach and notice that in general the only non-trivial constraint with NS-NS and $Q$-fluxes 
comes from the last equation of (\ref{Bianchi}) and is just
\eq{
\tilde{q}^{\Lambda A}\, h_{\Lambda} - q_{\Lambda}^A\, \tilde{h}^{\Lambda} = 0 \, .
}
Performing an S-duality transformation then leads to the generalized Bianchi identity 
\eq{
\label{bianchip}
\tilde{p}^{\Lambda A}\,\mathfrak{f}_{\Lambda} - p_{\Lambda}^A \,\tilde{\mathfrak{f}}^{\Lambda} = 0 \, .
}
Here we have used that both $(P,Q)$ and $(\mathfrak F,H)$ fluxes  transform as
an $SL(2,\mathbb Z)$  doublet.

\subsubsection*{Mass Scales}

Before turning to the uplift analysis in the next sections let us state our conventions and notation
for the different mass scales.
For the Planck mass we take $M_{\rm Pl} \sim 2.435 \cdot 10^{18}$ GeV, and for the
string mass $M_{\rm s} = ( \alpha ' )^{1/2}$. In terms of $M_{\rm Pl}$, the string and Kaluza-Klein
scales are given by
\eq{M_{\rm s} = \frac{\sqrt{\pi} M_{\rm {Pl}}}{s^{\frac{1}{4}} \mathcal{V}^{\frac{1}{2}}} , \quad 
M_{\rm KK} = \frac{M_{\rm Pl}}{\sqrt{4 \pi} \mathcal{V}^{\frac{1}{4}}} \, ,
}
where $s = e^{-\phi}$ and $\mathcal{V}$ is the volume of the Calabi-Yau manifold in Einstein frame in
string units. The moduli masses are determined by the eigenvalues of the canonically normalized mass matrix, 
which is defined as
\eq{
(M^2)^i_j = K^{ik} V_{k j} \, ,
}
where $V_{ij} = \frac{1}{2} \partial_i \partial_j V$. Finally, the gravitino mass reads
\eq{
M_{3/2}^2 = e^{K_0} |W_0|^2 \frac{M_{\rm Pl}^2}{4 \pi}
}
where $K_0$ and $W_0$ stand for the K\"ahler and superpotential evaluated at the minima.

\section{Uplifting to de Sitter}
\label{sec:uplift}

In this section we investigate whether, by adding additional positive
definite
contributions to the F-term scalar potential, one can directly find
scaling type, non-supersymmetric metastable minima that are
of de Sitter or Minkowski type. 

Recall that in the KKLT \cite{Kachru:2003aw} or LARGE volume scenario 
\cite{Balasubramanian:2005zx,Conlon:2005ki}, one starts with an
AdS minimum and adds the contributions of an $\overline{{\rm D}3}$-brane in a warped
throat. By varying the coefficient of this contribution, i.e. the
warp factor, one can continuously shift the cosmological constant in
the minimum from the negative AdS value to positive dS values.
In the first part of this section we analyze (in a concrete example)
the effect of adding an $\overline{{\rm D}3}$-brane to the F-term
flux-induced potential. 

In \eqref{potentialfull} we have recalled that for  $h^{2,1}_+>0$  the scalar
potential
receives an additional positive definite D-term contribution \eqref{dtermpot}.
Thus, it is tempting to try to uplift an AdS minimum by 
also turning on the fluxes contributing to this D-term.
We will analyze this question in the second part of this section.

\subsection{Uplift via $\overline{{\rm D}3}$-brane}
\label{s:up}

The common mechanism to uplift AdS vacua preserving stability 
is to introduce an $\overline{{\rm D}3}$-brane at a warped throat 
\cite{Kachru:2003aw,Kachru:2003sx}.
This generates a contribution to the scalar potential of the form 
\eq{
\label{vupgen}
V_{\rm up} = \frac{A}{ \mathcal{V}^{4\over 3}}\, {M^4_{\rm Pl}\over 4\pi}\, , 
}
with $A$ a positive constant depending on the warp factor in the throat.
Let us now  consider a concrete example showing what will happen
with a scaling type minimum after including the $\overline{{\rm D}3}$-brane
contribution to the scalar potential.

\subsubsection*{A stable AdS minimum}

Consider a CY manifold with $h^{11}_+=1$, $h^{11}_-=0$, $h^{21}_-=1$
and $h^{21}_+=0$.
Therefore, the total scalar potential after tadpole
cancellation is given just by the F-term. 
 The tree-level K\"ahler potential reads
\eq{
   K=-\log(S+\ov S) -3\log(T+\ov T) -3\log(U+\ov U)\,,
}
and the defining superpotential is given by
\eq{
\label{wpexdual}
W = - i f U + i h_0 S - 3 i h S U^2  - i q T \, .
}
According to \eqref{thebigW}, $\mff_1 = f$, $\tilde{h}^1 = - h$ and $q_0{}^1 = q$. 
In the following we will also denote $S=s + i c$, $T=\tau + i \rho$ and $U=v + iu$.

In absence of the $\overline{{\rm D}3}$-brane there is a 
completely stable supersymmetric AdS vacuum of scaling type. The
axionic moduli are fixed at
$\rho = c = u = 0$ , whereas the saxions are fixed at
\eq{
s = -\frac{5^{1/2}}{4} \frac{f}{\left( h h_0 \right)^{1/2}} \, , \quad v = \frac{5^{1/2}}{3} \left( \frac{h_0}{h} \right)^{1/2}   , \quad \tau = -\frac{5^{1/2} f}{2  q} \left( \frac{h_0}{h} \right)^{1/2} \, .
}
To be in the physical regime we choose fluxes such that 
\eq{
f < 0 \, , \qquad h_0 > 0 \, , \qquad  h > 0 \, , \qquad q > 0 \, .
}
To stay consistently in the perturbative regime,  one can choose  $|f| \gg 1$ and all
other fluxes ${\mathcal O}(1)$.
The value of the scalar potential at the minimum is given by
\eq{
V_0 = -\frac{9}{5^{5/2}\, 4} \frac{q^3 h^{5/2}}{f^2 h_0^{3/2}} \, \frac{M_{\rm Pl}^4}{4\pi} \, .
}
The normalized moduli masses are found to be
\eq{ \label{mm1_ads}
M^2_{\rm mod} =  \mu_i \frac{q^3 h^{5/2}}{f^2 h_0^{3/2}} \, \frac{M_{\rm Pl}^2}{4\pi}\, ,
}
with coefficients 
\eq{
\mu_i = \{ 0.4039, 0.2414, 0.1208; 0.5699, 0.1341,0.0442 \}\, .
}
The first three entries are saxionic while the last three are axionic. Thus, the lightest state is axionic.

\subsubsection*{Uplift to a Minkowski  minimum}

Now, we add the uplift term in \eqref{vupgen} for an $\ov{{\rm D}3}$-brane in the throat.
Searching directly for a stable Minkowski minimum with the axions kept
at the origin, one finds one, in which  the  saxions are shifted to
\eq{
\label{vevsaxq}
s = \frac{1}{3^{3/4}} \frac{f}{\left( h h_0 \right)^{1/2}} \, , \quad v = \frac{1}{3^{1/4}} \left( \frac{h_0}{h} \right)^{1/2} , \quad \tau = \frac{f}{3^{1/4} q} \left( \frac{h_0}{h} \right)^{1/2} \, .
} 
The warp dependent parameter $A$ is determined to be
\eq{
\label{Avalue}
A = \frac{3^{1/4}}{2}\frac{q h^{3/2}}{h_0^{1/2}}\, .
}
Clearly to have positive saxion vacuum expectation values in the
minimum,  the fluxes can be chosen in the regime
 \eq{
f > 0 \, , \qquad h_0 > 0 \, , \qquad  h > 0 \, , \qquad q > 0 \, .
} 
As a consequence, one gets  $A > 0$, as it should be.
Since the sign of $f$ is different from the supersymmetric AdS 
minimum, it is clear that this Minkowski vacuum is not literally a
continuous
uplift of the former, but constitutes a new non-supersymmetric, still
scaling type, Minkowski vacuum.

After the uplift, the normalized masses have the same flux dependence
(\ref{mm1_ads}) as in the AdS vacuum,  though the  numerical
coefficients change to 
\eq{
\mu_i = \{0.8034, 0.4868, 0.03942; 1.5559, 0.2116, 0.0811\}\, .
}
Observe that now the lightest state is a linear combination of
saxions. 

Utilizing the expressions given at the end of section \ref{sec:fluxscaling}, let us compute the other
relevant mass scales.
The gravitino mass has the same scaling behavior as (\ref{mm1_ads}) with coefficient $\mu_{3/2} = 0.3135$.
Moreover, the Kaluza-Klein and string scales are given by
\eq{
M_{\rm s}^2 = \frac{3^{3/4} \pi}{2^{3/2}} \frac{q^{3/2} h}{f^2 h_0^{1/2}} M_{\rm Pl}^2, \qquad 
M_{\rm KK}^2 = \frac{3^{1/2}}{16 \pi} \frac{q^2 h}{f^2 h_0} M_{\rm Pl}^2 \, 
}
so that the relevant ratios are determined as
\eq{
\frac{M_{\rm KK}^2}{M_{\rm s}^2} = \frac{1}{2^{5/2}3^{1/4}\pi^2} \left( \frac{q}{h_0}\right)^{1/2} \!\!\! ,\qquad 
\frac{M^2_{{\rm mod},i}}{M^2_{\rm KK}} = \frac{2^2 \mu_i}{3^{1/2}}\frac{q h^{3/2}}{h_0^{1/2}}.
}
Therefore, taking $h, q\sim {\mathcal O}(1)$ and $h_0\sim f\gg
1$ we get that parametrically the moduli are in their perturbative
regime and that  parametrically one can achieve the mass hierarchy 
$M_{\rm s} \parag M_{\rm KK}  \parag M_{\rm mod}$, which is important
for self-consistency of our approach.
Notice that for  $h_0 \gg 1$ we also obtain  $A \ll 1$.

Another  characteristic feature  of this model is that the fluxes do
not contribute to the ${\rm D}7$-brane tadpole whereas 
\eq{
\label{d3tada}
N_{{\rm D}3}= f\op h \, .
}
Notice that, while in the supersymmetric AdS vacuum $N_{{\rm D}3} <
0$,  in the 
Minkowski minimum $N_{{\rm D}3} > 0$. Increasing $f$ clearly gives a larger flux tadpole.

This example shows that adding an $\ov{{\rm D}3}$-brane to the
fluxed CY manifold the scalar potential admits  new  stable  scaling
type Minkowski vacua. Such vacua could serve as the starting point for
the realization of F-term axion monodromy inflation along the
lines proposed in \cite{Blumenhagen:2014nba,Blumenhagen:2015qda,Blumenhagen:2015kja}.

\subsubsection*{Uplift to a de Sitter minimum}

\begin{figure}[ht]
  \centering
  \includegraphics[width=0.7\textwidth]{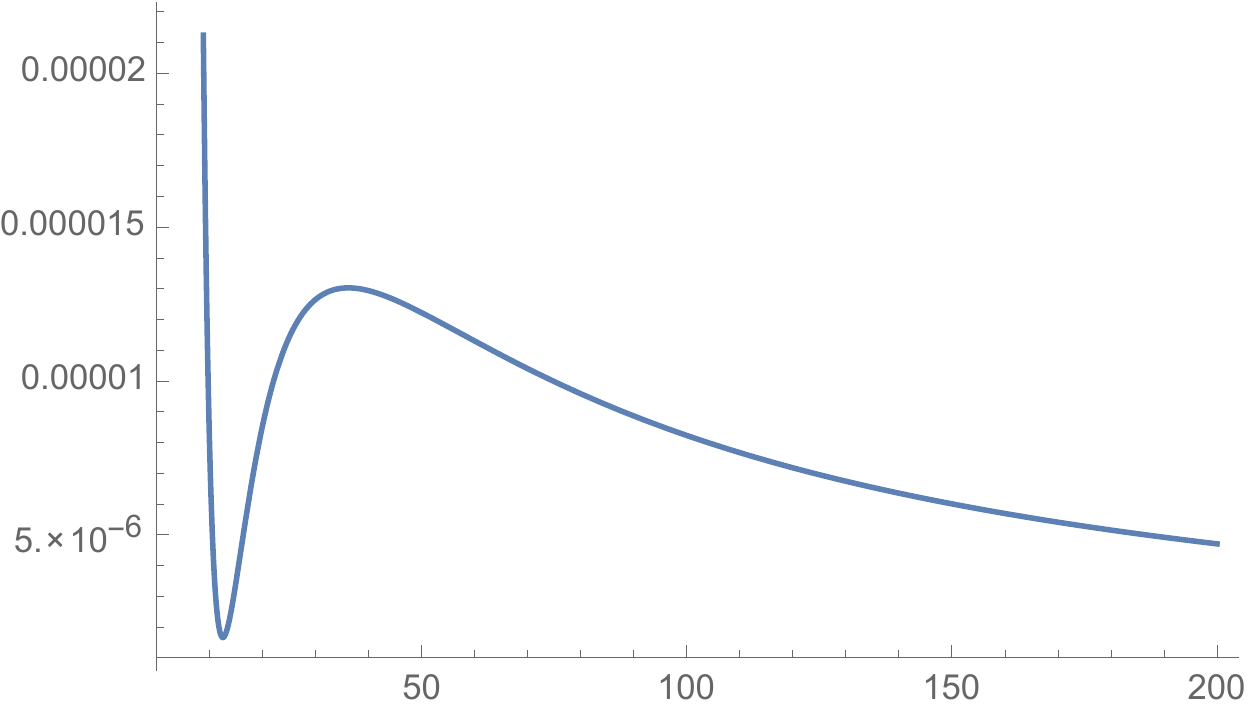}
  \begin{picture}(0,0)
  \put(-259,168){\footnotesize $V$}
   \put(-2,12){\footnotesize $\tau$}
  \end{picture}
  \caption{\small The scalar potential $V(\tau)$ in units of
    ${M_\text{Pl}^4 \over 4 \pi}$  for $\{s,v\}$ and the
    axions in their minimum. The fluxes are  $h_0=10$, $h=q=1$, $f=5$ and
    $A$ is chosen to give a de Sitter minimum.}
  \label{DeSitter}
\end{figure}
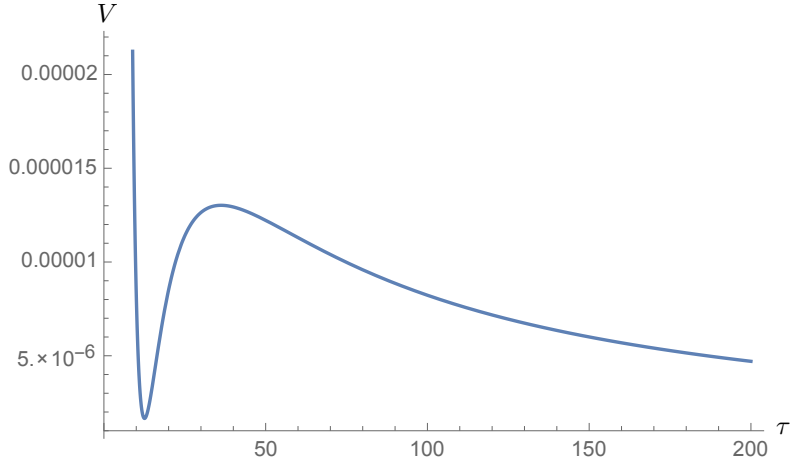

By choosing the parameter $A$ in the $\overline{{\rm D}3}$-brane potential
larger than  \eqref{Avalue}, one expects to also  get a de Sitter
vacuum.
Let us analyze this in an expansion in $\Lambda=V_0$, i.e. the value
of the scalar potential in the minimum.
Indeed changing the value of $A$, in the minimum, the axions are kept
at the origin while the saxions shift to
\eq{
\label{vevsaxq2}
s &= \frac{1}{3^{3/4}} \frac{f}{\left( h h_0 \right)^{1/2}} + \frac{2^4 \cdot 7}{3^{5/2}} \frac{f^3 h_0}{q^3 h^3} \Lambda + \mathcal{O}(\Lambda^2) \, , \\ 
v &= \frac{1}{3^{1/4}} \left( \frac{h_0}{h} \right)^{1/2} - \frac{2^4}{3^{2}} \frac{f^2 h_0^2}{q^3 h^3}\Lambda + \mathcal{O}(\Lambda^2) \, , \\
\tau &= \frac{f}{3^{1/4} q} \left( \frac{h_0}{h} \right)^{1/2} +\frac{2^4 \cdot 13}{3^2} \frac{f^3 h_0^2}{q^4 h^3} \Lambda + \mathcal{O}(\Lambda^2) \, .
} 
The parameter $A$ is determined to be
\eq{
A = \frac{3^{1/4}}{2}\frac{q h^{3/2}}{h_0^{1/2}} + \frac{2^2}{3^{1/2}}\frac{f^2 h_0}{q^2 h}\Lambda+ \mathcal{O}(\Lambda^2) \, .
}
In figure \ref{DeSitter} we display the form of the potential
for a choice of parameters leading to a de Sitter minimum.
Even though, for simplicity, only the dependence on a single  variable (here $\tau$) is
shown, the plot shows the expected behavior that is also familiar from
KKLT.
In particular, the dS minimum is only metastable as the potential
goes to zero for large $\tau$. 

\vspace{0.2cm}

The upshot is that for small $|\Lambda|$ one can continuously interpolate
from an AdS to a dS minimum. At certain critical values of $|\Lambda|$ the
vevs for the saxions in \eqref{vevsaxq2} can become negative and therefore unphysical. 
The normalized masses also get corrected at linear order in $\Lambda$ 
\eq{ 
M^2_{\rm mod} =  \left( \mu_i \frac{q^3 h^{5/2}}{f^2 h_0^{3/2}}  - \tilde{\mu}_i \Lambda  + \mathcal{O}(\Lambda^2)  \right) \, \frac{M_{\rm Pl}^2}{4\pi} \, ,
}
with coefficients 
\eq{
\mu_i = \{0.8034, 0.4868, 0.03942; 1.5559, 0.2116, 0.0811\}\, ,
}
and
\eq{
\tilde{\mu}_i = \{ 46.5221, 34.4038, 6.1852 ; 125.614, 6.5749, 3.6748 \}\,.
}
Note that the linear contribution of a positive cosmological constant 
decreases  the mass of all the moduli. Thus,
for too large $\Lambda$,
we expect the appearance of tachyonic states. 
The  Kaluza-Klein and string scale also receive corrections
so that the relevant ratios become
\eq{
\frac{M_{\rm KK}^2}{M_{\rm s}^2} &= \frac{1}{2^{5/2}3^{1/4}\pi^2} \left( \frac{q}{h_0}\right)^{1/2} -\frac{2^{3/2}}{3 \pi^2} \frac{f^2 h_0}{q^{5/2} h^{5/2}} \Lambda + \mathcal{O}(\Lambda^2) \, , \\ 
\frac{M^2_{\rm mod,i}}{M^2_{\rm KK}} &= \frac{2^2}{3^{1/2}}\mu_i \frac{q h^{3/2}}{h_0^{1/2}} + \frac{2^2}{3^3}\left( 2^5 \cdot 13 \cdot 3^{3/4} \mu_i + 3^{5/2}\tilde{\mu}_i \right)\frac{f^2 h_0}{q^2 h}\Lambda+ \mathcal{O}(\Lambda^2).
}
Thus we conclude that  the scaling behavior for all quantities is corrected at subleading order in $\Lambda$.

\subsection{D-term uplift}
\label{sec:dpot}

In this section we investigate a second possibility for uplift, namely
by taking the naturally appearing D-terms \eqref{dtermpot} into
account. These positive semi-definite contributions do only depend
on the saxionic modes and therefore  do not change the axion stabilization.
For concreteness, we consider a model with Hodge numbers
$h^{2,1}_+=1$,  $h^{2,1}_- =1$,  $h^{1,1}_+ =1$ and $h^{1,1}_- =0$.
The  derivation of the explicit form of the corresponding D-term
potential is presented  in some detail in Appendix A. The final result is
\eq{
\label{dpotbp}
V_D= \frac{\delta}{v \tau^2} \left(g - \frac{ r \op \tau}{ 3 \op s}\right)^{\!\! 2} \, ,
}
where $r=f_{\hat 1\, 0}$, $g=f_{\hat 1\, 1}$, and
$\delta$ is  an unphysical  positive constant which can be absorbed in a
redefinition of the fluxes.
The superpotential leading to an additional F-term potential $V_F$
is chosen to be
\eq{\label{supm4}
W = i \mff U +i \tilde{\mff}  U^3 -i {h} S +i {q} T  \, ,}
where we redefined $\mff_1 = -\mff$, $\tilde{\mff}^0 = \tilde{\mff}$,
${h}_0 = -{h}$ and ${q}_0{}^{1} = -{q}$.
After imposing the Bianchi identities \eqref{biadd},  the D-term becomes
\eq{ \label{VdM4}
V_D  = \frac{\delta g^2}{\tau^2 v} \left( 1 + \frac{{q}}{3 {h}}\frac{\tau}{s} \right)^{\! 2} \, .
}
The total scalar potential $V=V_F+V_D$, by a suitable choice of
$\delta$,
admits  a tachyon-free (stable) Minkowski minimum with axions fixed at
\eq{
\label{axionDup}Re: 
\Theta = {q} \rho - {h} c = 0 ,\quad u = 0 \, , } 
and saxions at
\eq{
\label{saxionDup}
s = \gamma_1 \frac{{\mff}^{3/2}}{h\tilde\mff^{1/2} } ,\quad \tau = \gamma_2 \frac{{\mff}^{3/2}}{{q} \tilde\mff^{1/2}} ,\quad 
v = \gamma_3 \left( \frac{\mff}{\tilde{\mff}} \right)^{1/2} \, ,}
while the constant $\delta$ is given by
\eq{
\label{deltaM4}
\delta g^2 = \gamma_4 \frac{ h{q}\op \tilde\mff }{{\mff}} \, .
}
The numerical coefficients above are
\eq{ \gamma_i = \{ 0.1545, 1.5761 , 1.0318, 0.0044\} \, .}
We can stay in the physical region, and have $\delta > 0$, by choosing 
 $\mff, \tilde{\mff}, {h}, {q} > 0$. The saxions are 
 fixed in their perturbative regime for $\mff \gg \tilde{\mff}$
 and $\tilde{\mff} ,h,q$ of order one.
The normalized masses are given by
\eq{
M_{\rm mod,i}^{2} = \mu_i \frac{h {q}^3\, \tilde\mff^{5/2}}{\mff^{9/2}} \frac{M_{\rm Pl}^2}{4 \pi} \, ,
}
with prefactors
\eq{\mu_i = 
\{0.6986, 0.0152, 0.1318; 0.2594 , 0.0524, 0\}\, .
}
Therefore, as expected there is one massless axion and the next
lightest state is a saxion.
The KK and string scales  are given by
\eq{
M^2_{\rm s} = 1.428 \frac{ h^{1/2} \,q^{3/2} \,\tilde\mff }{ \mff^3 } M_{\rm Pl}^2 ,\quad  M^2_{\rm KK} = 0.008 \frac{{q}^2\, \tilde\mff}{{\mff}^3} M_{\rm Pl}^2 \, .
}
The ratio of the KK and string scale is
\eq{
\frac{M^2_{\rm s}}{M^{2}_{\rm KK}} = 178
\frac{{h}^{1/2}}{{q}^{1/2}} \, ,
\qquad 
 \frac{M^2_{\rm KK}}{M^{2}_{\rm mod}} = { 0.1\over \mu_i}{1\over  h  q} \frac{\mff^{3/2}}{\tilde\mff^{3/2}}\,.
}
We can guarantee that $M_{\rm s}>M_{\rm KK}$ for $ h> q$ and $M_{\rm
  KK}>M_{\rm mod}$ for  $\mff \gg \tilde{\mff}$. Therefore, in the
perturbative regime the 
KK scale is parametrically heavier than the moduli mass  scale.
Since we have in addition one massless axion, this model is a good
starting point for realizing F-term axion monodromy inflation.

\section{Axion monodromy inflation}
\label{sec:ami}

In this section we study the  inflaton potentials resulting from the Minkowski models obtained by 
including the D-term generated by non-geometric fluxes. 
One important difference to the analysis in \cite{Blumenhagen:2015kja, Blumenhagen:2015qda}
is that now the uplift to zero or positive cosmological constant is
not done by hand.
Recall that to guarantee the consistency of the effective field theory
approach as well as to realize a model of single field inflation,
one has to stabilize the moduli such that the following hierarchy of mass
scales is realized 
\eq{
\label{hierarchy1}
M_{\rm s} > M_{\rm KK} > M_{\rm inf}> M_{\rm mod} > H_{\rm inf} > M_{\rm \theta} \, ,
}
where $H_{\rm inf}$ is the Hubble scale during inflation and $M_{\rm
  inf}=V^{1\over 4}_{\rm inf}$ the mass scale of inflation. Assuming a
constant uplift, it was demonstrated in \cite{Blumenhagen:2015kja}, how difficult
it is to obtain such a hierarchy.

\subsection{Effective field theory approach}
\label{ss:analytical}

For the model in section \ref{sec:dpot} with the D-term uplift,  we have one unstabilized and therefore massless axion. According
to \cite{Blumenhagen:2015kja, Blumenhagen:2015qda} we can try 
to generate a parametrically small mass for this axion by turning
on additional fluxes and scale the former fluxes by a parameter
$\lambda$. A good candidate for the extra flux is a $P$-flux
\cite{Aldazabal:2006up} so that  we now take the extended superpotential 
\eq{
      W=\lambda W_0 -i  p\, S\,T\, U\, ,
}
where $W_0$ is given in \eqref{supm4}.
Note that the full set of fluxes in $W$ is not constrained by Bianchi identities.
The new superpotential generates an F-term scalar potential in which the former
terms scale with $\lambda^2$. In the large $\lambda$ limit we would
like to get the old minimum. To this end we scale the D-term potential
as
\eq{
   V_D  = \lambda^2\frac{(\delta_0+\Delta\delta) g^2}{\tau^2 v} \left( 1 + \frac{{q}}{3 {h}}\frac{\tau}{s} \right)^{\! 2} \, .
} 
Here we have split $\delta$ into 
$\delta_0$ given by the former value \eqref{deltaM4} plus a correction term
$\Delta\delta$ needed to guarantee a Minkowski minimum also after
including the $P$-flux. 

We will assume that $\lambda$ is large and
work in a $1/\lambda$ expansion. The leading order contribution to the shift 
in the uplift parameter turns out to be
\eq{
           \Delta\delta\sim - {p\, \mff\over \lambda \, g^2 }\, .
}
Assuming $\lambda$ sufficiently
large we can also integrate out the heavy moduli and derive an 
effective potential for the former massless axion which is the orthogonal combination to $\Theta$ in \eqref{axionDup}.
Since at the minimum $\Theta=0$ we can take this axion to be  $\theta=c$.
Integrating out the heavy moduli we obtain
the effective quartic potential 
\eq{
        V_{\rm eff} =B_{1}\, \theta^2 + B_2\,\theta^4
}
with
\eq{
    B_1\sim {\lambda\, p \, h^2\, q^2 \,\tilde\mff^{5/2}\over \mff^{11/2}}\,,\qquad
    B_2\sim {p^2\, h^3 \, q\, \tilde\mff^{5/2}\over \, \mff^{13/2}}\,.
}
For sufficiently large $\lambda$, one can ensure that the quadratic
term is dominant for say $\theta$ of ${\mathcal O}(10)$, as needed for 
large field inflation.

After canonical normalization, we can compute the mass of the
inflaton. For the ratios of mass scales we find
\eq{
      {M^2_{\rm KK}\over M^2_{\rm mod}}\sim {\mff^{3/2}\over
          \lambda^2\, h\,q \,\tilde\mff^{3/2}}\,,\qquad
     {M^2_{\rm mod}\over M^2_{\theta}}\sim {\lambda\, h\, q\, \tilde\mff\over
          p \, \mff^{2}}\,.
}
Indeed, for large $\lambda$ the inflaton mass becomes parametrically
lighter than the mass of all the other moduli, which however are in
danger of becoming heavier than the KK scale.
Taking the product of the two mass ratios one gets
\eq{
    {M^2_{\rm KK}\over M^2_{\rm mod}}\, {M^2_{\rm mod}\over
      M^2_{\theta}}\sim  {1\over \lambda\,  p\, \mff^{1/2}\,
      \tilde\mff^{1/2}\,. }
}
Clearly, as long as all these fluxes are positive integers and
$\lambda$ large, it is in principle impossible to have both 
mass ratios  larger than one, as desired.
Note that this problem was already encountered in \cite{Blumenhagen:2015kja}.

One potential loophole in this no-go result is the assumption that
all fluxes are integer quantized. 
In fact, as also realized in \cite{Blumenhagen:2014nba}, 
the prepotential for the complex structure moduli in the large complex
structure limit is subject to perturbative and non-perturbative corrections,
which take the general form (see for instance
\cite{Hosono:1994av})\,\footnote{Note that the terminology of
  perturbative and non-perturbative corrections is actually taken from the mirror
dual side, where the complex structure moduli  are exchanged with the
K\"ahler moduli.}
\eq{
  \label{f_corr}
  \widetilde F = F + \frac{1}{2} \op a_{ij} X^i X^j + b_i X^i X^0 + \frac{1}{2}\op i\op\gamma \bigl( X^0\bigr)^2
  + F_{\rm inst.} \,,
}
with the usual cubic term ${F}={1\over 6}d_{ijk} X^i X^j X^k/X^0$.
Here, the constants $a_{ij}$ and $b_i$ are rational  numbers, while
$\gamma$ is real. 
From the point of view of the mirror dual threefold $\hat M$, they are determined as 
\eq{
a_{ij}&=-\frac{1}{2}\int_{\hat M} \hat\omega_i \wedge \hat\omega_j
  \wedge \hat \omega_j\,,\quad 
b_i=\frac{1}{24}\int_{M}c_2(\hat M) \wedge \hat\omega_i,\quad {\rm
  mod}\ \mathbb Z\\
i\,\gamma&=\frac{1}{(2 \pi i)^3} \chi(\hat M) \zeta(3)\,,
}
with the second Chern class $c_2(\hat M)$, the Euler number of the
internal space $\chi(\hat M)$ and a basis of harmonic $(1,1)$-forms $\hat\omega_i$.
These constants can be smaller than one, but not arbitrarily small.
Note that when evaluating  the superpotential \eqref{thebigW}, 
the  corrections $a_{ij}$ and $b_i$ can  be incorporated by the
following shifts in the fluxes
$g_\Lambda\in\{\mff_\Lambda, f_{\Lambda a}, q_\Lambda{}^\alpha\}$
\eq{
\label{fluxshift}
     g_0 = g_0 -b_i\op \tilde g^i  \,,  \qquad g_i = g_i -a_{ij}
     \op\tilde g^j - b_i \op\tilde g^0 \,.
}
Recall that the purely imaginary contribution $i\op\gamma$ corresponds
to  $\alpha'$-corrections to the  K\"ahler potential for the 
K\"ahler moduli in a mirror-dual setting. In the large
complex-structure regime we are 
employing here, these corrections can be neglected. 
Similarly, in this regime also the 
non-perturbative corrections $F_{\rm inst.}$ are negligible. 
To summarize,  the polynomial  corrections to the prepotential can be
incorporated by a rational shift in the fluxes. This at least
motivates the numerical approach to be adopted in the following
section \footnote{Let us mention that in other recent works \cite{Kallosh:2014oja,Blaback:2015zra} on de
 Sitter vacua of string theory, the fluxes were also chosen to be rational.}.

\subsection{Numerical analysis of inflation}
\label{ss:numerical}

\begin{figure}[!h]
 \centering
 a)\includegraphics[width=0.45\textwidth]{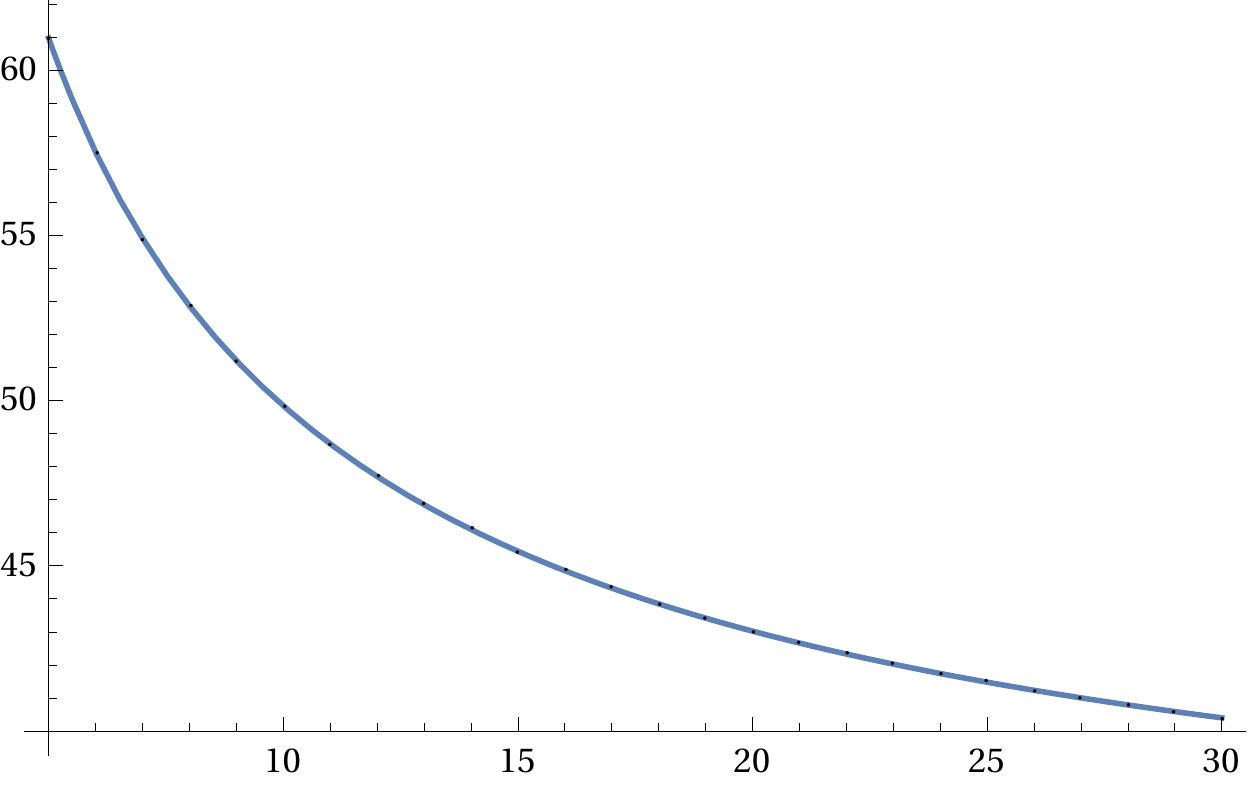}
 \begin{picture}(0,0)
\put(-180,120){\footnotesize $\frac{M^2_{\rm s}}{M^2_{\rm KK}}$}
   \put(-5,10){\footnotesize $\lambda$}
  \end{picture}
 b)\includegraphics[width=0.45\textwidth]{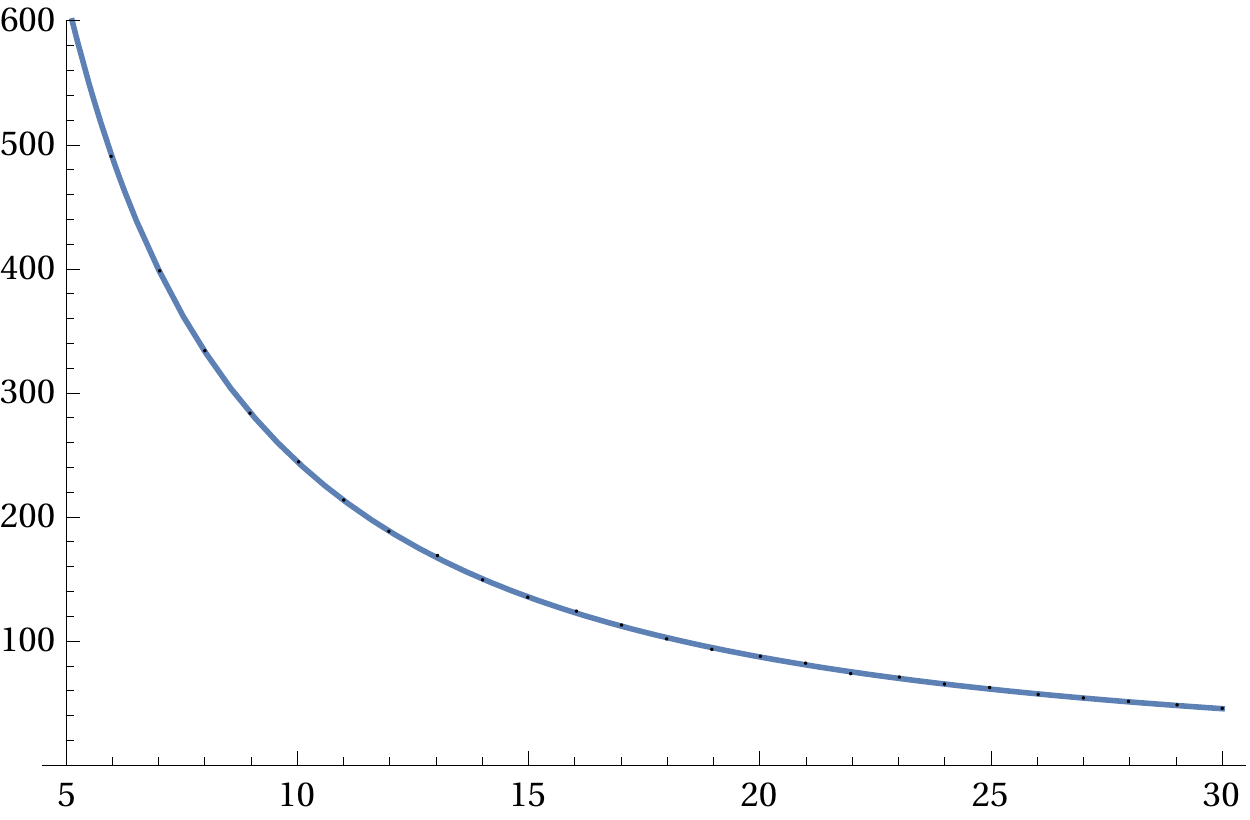}
 \begin{picture}(0,0)
  \put(-175,120){\footnotesize $\frac{M^2_{\rm KK}}{M^2_{\rm mod}}$}
   \put(-5,10){\footnotesize $\lambda$}
  \end{picture}
\caption{\label{fig:Mratio} Ratio of relevant mass scales for a) string scale over Kaluza-Klein scale
and b) the Kaluza-Klein scale over the
heaviest modulus. Fluxes are chosen rational with values 
$h = 1/220$,  $\tilde{\mff} = 1/1810$, $\mff = 6/49$, $q = 1/8$, $g = 1/10$ and $p = 1/10000$.}
\end{figure}
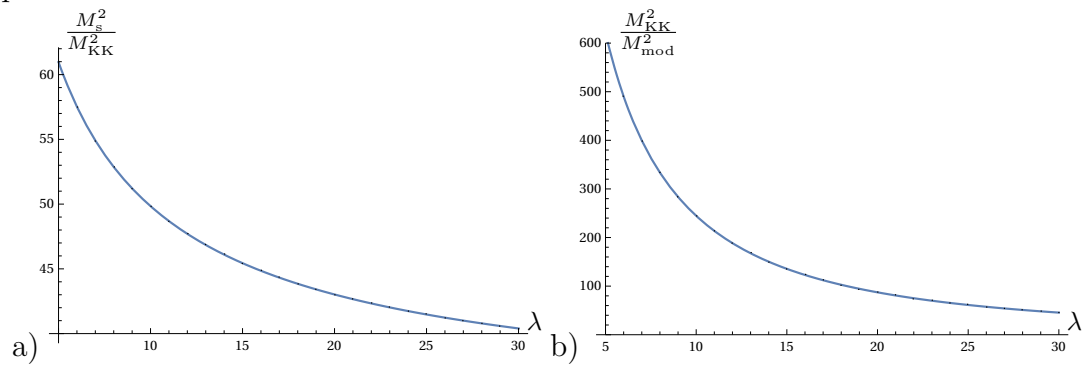

Instead of  pursuing an effective approach, as in the previous
subsection, we now follow an exact, though numerical, approach to analyze the same model.
In practice we choose initial (phenomenologically motivated) values of the fluxes, compute
the exact scalar potentials in terms of all moduli fields and
then look numerically for stable Minkowski minima.
We are particularly interested in determining whether there exists
a choice of (rational) fluxes so that we can concretely realize the
hierarchy of mass scales shown in  \eqref{hierarchy}.
In figure \ref{fig:Mratio} we display, for a certain choice of fluxes, 
the behavior of some relevant mass ratios as the scaling parameter
$\lambda$ is varied. 

\begin{figure}[!ht]
 \centering
a)\includegraphics[width=0.45\textwidth]{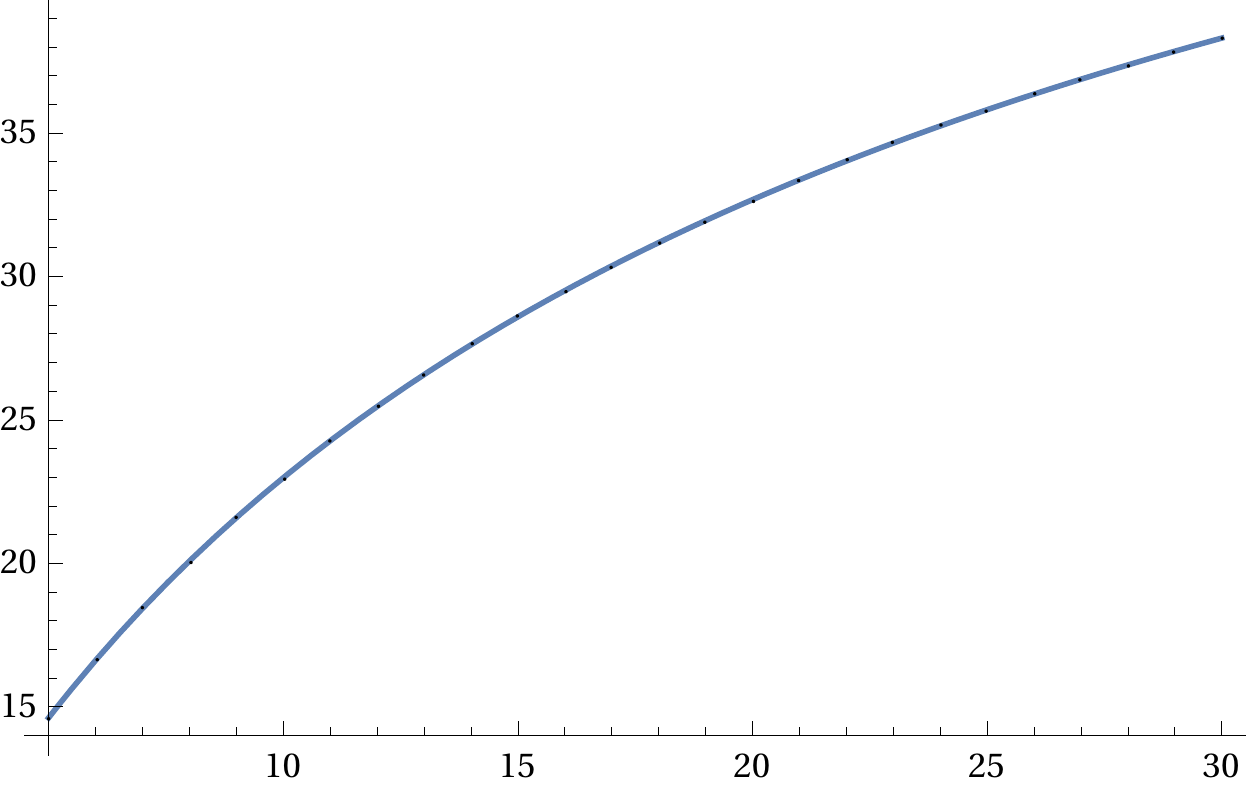}
 \begin{picture}(0,0)
  \put(-180,118){\footnotesize vev}
  \put(-80,95){\footnotesize $\langle s \rangle$}
   \put(-5,10){\footnotesize $\lambda$}
  \end{picture}
b)\includegraphics[width=0.45\textwidth]{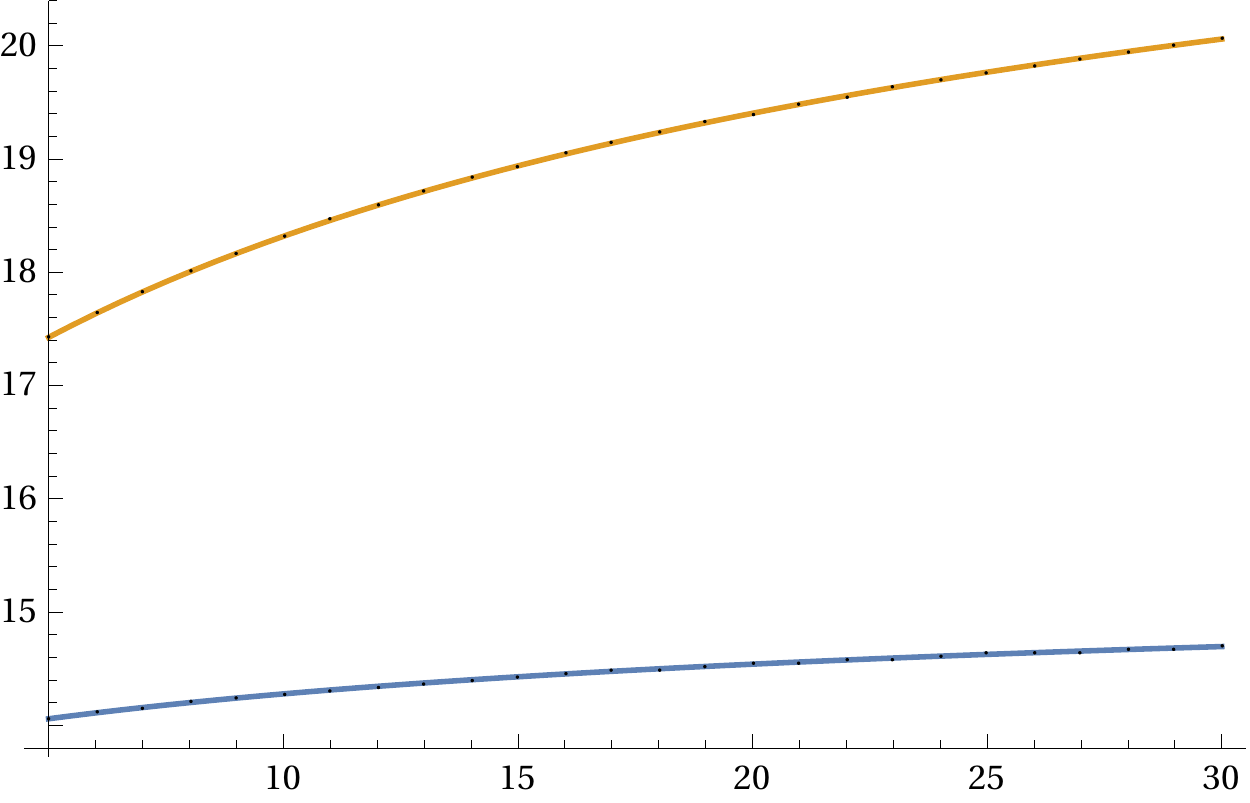} 
\begin{picture}(0,0)
  \put(-180,118){\footnotesize vev}
  \put(-80,110){\footnotesize $\langle \tau \rangle$}
  \put(-80,25){\footnotesize $\langle v \rangle$}
   \put(-5,10){\footnotesize $\lambda$}
  \end{picture}
\caption{\label{fig:vev-s} Vev's of  the saxionic moduli for a) $s$ and b) $\tau$ and $v$.}
\end{figure}
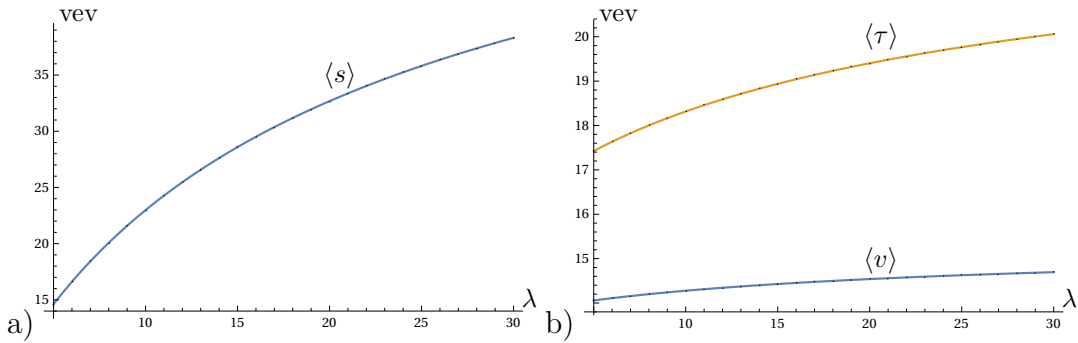

From figure \ref{fig:Mratio} we conclude that for all values of $\lambda$ the KK and string mass are separated by a factor
of $\mathcal{O}(10)$. Moreover, the heaviest moduli mass is lower than
the KK scale by a factor of $\mathcal{O}(10^2)$ for small  $\lambda$ whereas
even for values of  $\lambda\sim 30$, the heaviest moduli mass is lower than the KK scale by a factor of $\mathcal{O}(10)$. 
Thus, we have control over these scales with hierarchy
\eq{
M_{\rm Pl} > M_{\rm s} > M_{\rm KK} > M_{\rm mod}\,.
}
The axions are fixed at
\eq{
\Theta  = \theta = u = 0 \, , 
}
whereas the saxions vary with $\lambda$ as shown  in figure
\ref{fig:vev-s} for the same fluxes as in figure \ref{fig:Mratio} 

We observe that as $\lambda$ increases the saxionic vevs
increase so that  we can trust the perturbative expansion for all
$\lambda$.
Let us mention that  for $\lambda < 5$  tachyons appear in the
spectrum that are not shown in figure \ref{fig:Mratio}.
Finally, for all $\lambda$ the lightest state is related to
the axion $c$ and its mass is smaller than the next heavier state by a
factor of $\mathcal{O}(10^2)$. In the following will consider $c$ as the inflaton candidate.

\begin{figure}[!h]
 \centering
 \includegraphics[height=5cm]{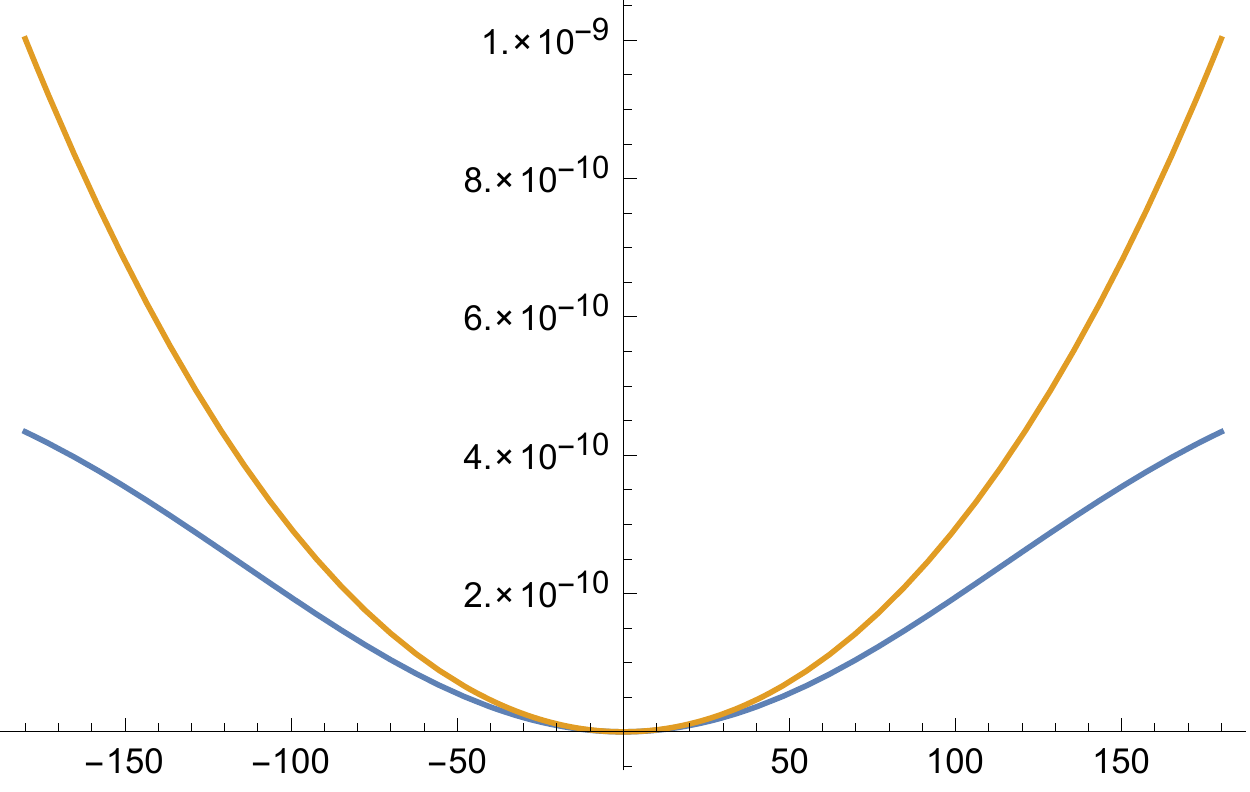}
 \begin{picture}(0,0)
  \put(-113,140){\footnotesize $V$}
   \put(0,10){\footnotesize $\theta$}
  \end{picture}
\caption{\label{fig:Vback1} Backreacted (blue line) and quadratic potential in units $\frac{M_{\rm Pl}^4}{4\pi}$ 
given by eq.~\eqref{vback2} for $h = 1/220$,  $\tilde{\mff} = 1/1810$, $\mff = 6/49$, $q = 1/8$, $g = 1/10$, $p = 1/10000$ and $\lambda = 10$.}
\end{figure}
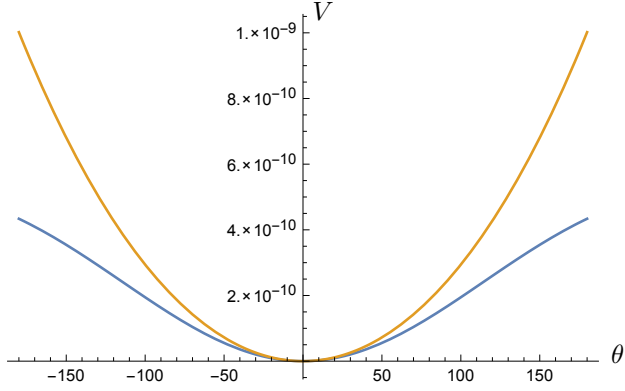

Next, for the values of fluxes shown above and choosing $\lambda=10$, 
we  consider the backreaction effect \cite{Dong:2010in} 
of the slowly rolling light axion $\theta=c$. 
The main task is to solve the extremum conditions $\partial_i V=0$ to obtain the saxions as functions of $\theta$.
Due to the complexity of the scalar potential we can only perform a
numerical analysis. Fixing all the heavy moduli at the minimum, the
effective scalar potential turns out to be 
\eq{
\label{vback2}
V_{\rm eff}(\theta) \approx B_1\, \theta^2 + B_2\, \theta^4 \, ,
}
where $B \cdot 10^{14} = \{2.8711 , 6.8314\cdot10^{-6} \}$. 
Thus, the quartic term is suppressed by a factor of $\mathcal{O}(10^{-6})$,
and the effective scalar potential for sufficiently small $\theta$ has a quadratic behavior.
To have a Minkowski vacuum it must be $\delta \cdot 10^{7} = 6.0647$.
Figure \ref{fig:Vback1} shows the scalar potential including the backreaction, together with the effective
scalar potential given in \eqref{vback2}. From figure \ref{fig:Vback1}, we observe that near  $c = 0$
both potentials match, while the backreaction modifies the shape of the scalar potential
for larger values of the inflaton $\theta$, producing a plateau-like behavior.

In order to compute the cosmological quantities $n_s$, $\epsilon$, $\eta$ and $N_e$, we first calculate the slow-roll parameters $\epsilon$ and $\eta$ as in \cite{Blumenhagen:2015qda}. 
Recall that for the Lagrangian ${\mathcal L} =\frac12 f(\theta)^2 (\partial \theta)^2 + V(\theta)$ the slow-roll parameters are given by
\begin{align}
\epsilon &= \frac{1}{2 f} \left( \frac{\partial V}{V} \right)^2\, ,  \qquad \eta = \frac{\partial^2  V}{f V}-\frac{\partial f \partial V}{2 f^2 V}.
\end{align}
The end of inflation is determined by the point on the moduli space in which the slow-roll conditions
are violated, i.e. $\epsilon \sim 1$. 
The starting point for the inflationary trajectory is chosen in such a way that $n_s = 0.9667 \pm 0.004$ \cite{Ade:2015lrj}. 
The e-foldings
as well as the tensor-to-scalar ratio are then derived from
\eq{
r = 16 \epsilon_{*} ,\quad N_e = \int_{\theta_{\rm end}}^{\theta^{*}}d\theta \frac{f V}{\partial V} \, ,
}
evaluated at the pivot scale $\theta^*$.

The value of the amplitude of the scalar power spectrum reported experimentally is 
$\mathcal{P} = (2.142\pm 0.049)\cdot 10^{-9}$, and it 
is determined from the Hubble scale and $\epsilon$ at the pivot scale by
\eq{
\mathcal{P} \sim \frac{H^{2}_{\rm inf}}{8 \pi^2 \epsilon_{*} M_{\rm Pl}^2} \, .
}
From this expression one derives  the Hubble scale during inflation. For the choice of fluxes 
mentioned above, we get the inflationary parameters
in table \ref{tab:param}.
\begin {table}[htbp]
\begin{center}
\begin{tabular}{ |c | c |c |c |  }
  \hline
 Parameter & Value \\
 \hline
 $\Delta c$         & $93 \, M_{\rm Pl}$ \\
 $N_e$                     & $61$        \\
 $r$                    & $0.0980$             \\
 $n_s$                    & $0.9667$        \\
 $\mathcal{P}$        & $2.14 \cdot 10^{-9}  $    \\
 $M_{\rm s}$            &  $1.04 \cdot 10^{17} \, $GeV \\
 $M_{\rm KK}$            & $1.49 \cdot 10^{16} \, $GeV \\
 $M_{\rm inf}$            & $ 4.89 \cdot 10^{15}\,  $GeV  \\
 $M_{\rm mod}$            & $\{11.99,4.81,2.38,6.81,2.47\} \cdot 10^{14}\,  $GeV  \\
 $H_{\rm inf}$                & $7.82 \cdot 10^{13} \, $GeV        \\
 $M_{\theta}$            &  $1.70  \cdot 10^{13}\,  $GeV        \\
\hline
\end{tabular}
\caption{\label{tab:param} Summary of inflationary parameters for $\lambda = 10$.}
\end{center}
\end{table}
Thus, for $9.44 < \theta < 104$ one collects $60$-efoldings for the reported spectral index $n_s$. From
table \ref{tab:param} we obtain the hierarchy of mass scales
\eq{
M_{\rm Pl} > M_{\rm s} > M_{\rm KK} > M_{\rm inf} > M_{\rm mod} > H_{\rm inf} > M_{\theta} \, 
}
with all individual scales showing the expected value. The value for the
tensor-to-scalar ratio lies on the boundary of being ruled out
experimentally and is a bit smaller than the value for quadratic
inflation. For the same model, in Appendix \ref{sec:infex} we consider
a different value of $\lambda$ leading to a lower value of $r$.

This numerical example shows that by allowing rational values of the
fluxes,  in particular those smaller than one, it is in principle
possible to freeze all moduli such that the above desired hierarchy of
mass scales is realized. Of course for a concrete Calabi-Yau manifold
the parameters for the polynomial terms in the prepotential \eqref{f_corr}
are fixed and therefore the admissible fluxes are more constrained
than assumed in our  phenomenological study. In particular, 
non-vanishing  fluxes could not be smaller than $|1/24|$ and 
the flux $\tilde \mff$ according to \eqref{fluxshift} would still be an
integer.

\section{Conclusions}
\label{sec:con}

The previous work \cite{Blumenhagen:2015kja} proposed a scheme of high
scale moduli stabilization, designed to realize
axion monodromy inflation. All minima discussed there were of AdS type 
and thus had  a negative cosmological constant.
The main aim of this paper was to build more realistic models by 
identifying working uplift mechanisms.
We considered  two possible energy sources  contributing a positive semi-definite term to
the scalar potential, namely an $\ov{{\rm D}3}$-brane 
or a D-term induced by geometric and non-geometric fluxes for non-zero $h^{2,1}_+$.
Both approaches did not uplift initial  flux-scaling minima, but
rather led to new  de Sitter and Minkowski minima still of
flux-scaling type. 
 
We explored to what extent the uplifted models could serve as starting points for the realization
of axion monodromy inflation with a parametrically controlled
hierarchy of induced mass scales.  
As in the previous study, we found that
the required hierarchy among the KK scale, the moduli mass scale and the
axion mass scale was not achieved as long as  we insisted  on integer
fluxes. Recalling that the perturbative corrections to the prepotential
of the complex structure moduli effectively lead to a redefinition
of the fluxes, we performed  a numerical model search admitting
also rational values of all fluxes. In this way we pinpointed two examples where all the desired
properties could be fulfilled.

This last result should be considered as an interesting observation.
Clearly, we are still far from a fully fledged
string theory model. A concrete
Calabi-Yau manifold with an orientifold projection has not been specified. Moreover, it has not
been established conclusively that the considered vacua of four-dimensional
gauged supergravity do uplift to full solutions of ten-dimensional string theory.

\vspace{0.7cm}

\emph{Acknowledgments:}
We would like to thank M. Fuchs and A. Guarino for valuable comments. We are specially grateful to E. Plauschinn for very useful
discussions and a careful reading of appendix A. 
A.F. thanks the Max-Planck-Institut f\"ur Physik as well as the Ludwig-Maximillians-Universit\"at for hospitality
and support during the early stages of this work. The research of
C.D. is supported by CONACyT through Assistantship No.
237801. C.D. also wants to thank the Max-Planck-Institut f\"ur Physik for its 
kind hospitality.

\newpage

\appendix

\section{D-term potential from $h^{2,1}_+ $ vector multiplets}

As we have seen in section \ref{sec:dpot}, the D-term potential \eqref{dtermpot} can be used to uplift
the cosmological constant to zero. In this appendix we will discuss the form of this D-term in some detail. 

We focus on the case $h^{2,1}_+=1$, and $h^{2,1}_- =1$. To simplify we also take
$h^{1,1}_+ =1$ and $h^{1,1}_- =0$. In the notation of section \ref{sec:fluxscaling} we turn on the fluxes
\eq{
\label{fch}
f_{\hat 1\, 0}=r \, , \qquad f_{\hat 1\, 1}=g \, 
}
whereas $\tilde f_{\hat 1\, 0}=0$ and  $\tilde f_{\hat 1\, 1}=0$. The D-term potential is then given by
\eq{
\label{dpot1}
V_D =-\frac{M^4_{\rm Pl}}{2} \op \frac{D_{\hat 1}^2}{ {\rm Im}\, \mathcal N} \, ,
}
where $\mathcal N = {\mathcal N}_{\hat 1 \hat 1}$ will be determined shortly, and $D_{\hat 1}$ reads
\eq{
\label{dhat1}
D_{\hat 1} = \frac{g t}{\mathcal V} - r  e^\phi = \frac{3}{\tau} \left(g - \frac{r\tau}{3 s}\right) \, .
}
Here we have used ${\mathcal V}=\frac16 \kappa t^3$, $T+\ov{T}=\kappa t^2=2\tau$, and  $s = e^{-\phi}$. 

Let us now compute the remaining ingredient ${\rm Im}\, \mathcal N$. As explained in \cite{Grimm:2004uq}, when 
properties of the orientifold projection are taken into account, the relation between the relevant period matrix elements 
and the prepotential reduces to
\eq{
\label{per1}
{\mathcal N}_{\hat \lambda \hat \sigma} = \ov{F}_{\hat \lambda \hat \sigma} \, .
}
In the right hand side the complex structure deformations associated to $h^{2,1}_+$ are set to zero. 
Working in the large complex structure limit the prepotential in our case can be expressed as
\eq{
\label{pF}
F=\frac{1}{X^0} \left(d_{111} X^3 + 3 d_{1\hat 1 \hat 1} X Z^2 \right) \, ,
}
where $X=X^1$ and $Z=X^{\hat 1}$. 
The form of the cubic prepotential follows imposing that under the orientifold involution $X$ and $X^0$ are even,
whereas $Z$ is odd.
The complex structure parameter associated to $h^{2,1}_-=1$ is defined as
\eq{
\label{udef}
U = -i\frac{X}{X^0} = v + i u\, .
}
We then find
\eq{
\label{per2}
{\rm Im}\, \mathcal N = -3  d_{1\hat 1 \hat 1} \left(U+\ov{U}\right) = - 6 d_{1\hat 1 \hat 1} \op v\, .
} 
Recall also that the K\"ahler potential for the complex structure sector is given by
\mbox{$K_{\rm cs} = -\log\left(- i \int_{\mathcal X} \Omega\wedge \ov\Omega \right)$}.
In our model we obtain $K_{\rm cs} = -3\log\left(U+\ov{U}\right)$,
setting $X^0=1$ and $d_{111}=1$. Thus, in physical regime $v > 0$. Now, since
the D-term potential \eqref{dpot1} must be positive definite, ${\rm Im}\, \mathcal N < 0$. Therefore,
$d_{1\hat 1\hat 1} > 0$. 

Substituting various preceding results in \eqref{dpot1} gives the  D-term potential
\eq{
\label{dpot3}
V_D= \frac{\delta}{v \tau^2} \left(g - \frac{ r \op \tau}{ 3 \op s}\right)^{\!\! 2} \, ,
}
where $\delta$ is a positive constant. Observe that this potential depends on all
the saxions in the model. The fluxes entering in $V_D$ are related to the action of the twisted differential
${\mathcal D}$ on the even $(2,1)$ forms. Such fluxes do not enter at all in the superpotential $W$ that determines
the F-term potential. However, there are Bianchi identities that mix $r_{\hat \lambda}$ and $f_{\hat \lambda \alpha}$
with NS-NS and $Q$-fluxes that might appear in $W$. In the model at hand the mixed BI constraints are
\eq{
\label{biadd}
r\, \tilde h^{\lambda} + g\, {\tilde q}^{\lambda\, 1} = 0 \, , \qquad r\, h_{\lambda} + g\, q_{\lambda}{}^1 = 0 \, ,
}
for $\lambda=0,1$.

\section{A second example of axion inflation}
\label{sec:infex}

Let us consider the same model as in section 4 but choose the limit case with $\lambda = 5$. Recall that for $\lambda < 5$,
tachyons appears on the spectrum. As in the previous case, the lightest state is axionic and related
to $\theta = c$.

\begin{figure}[!h]
 \centering
 \includegraphics[height=5cm]{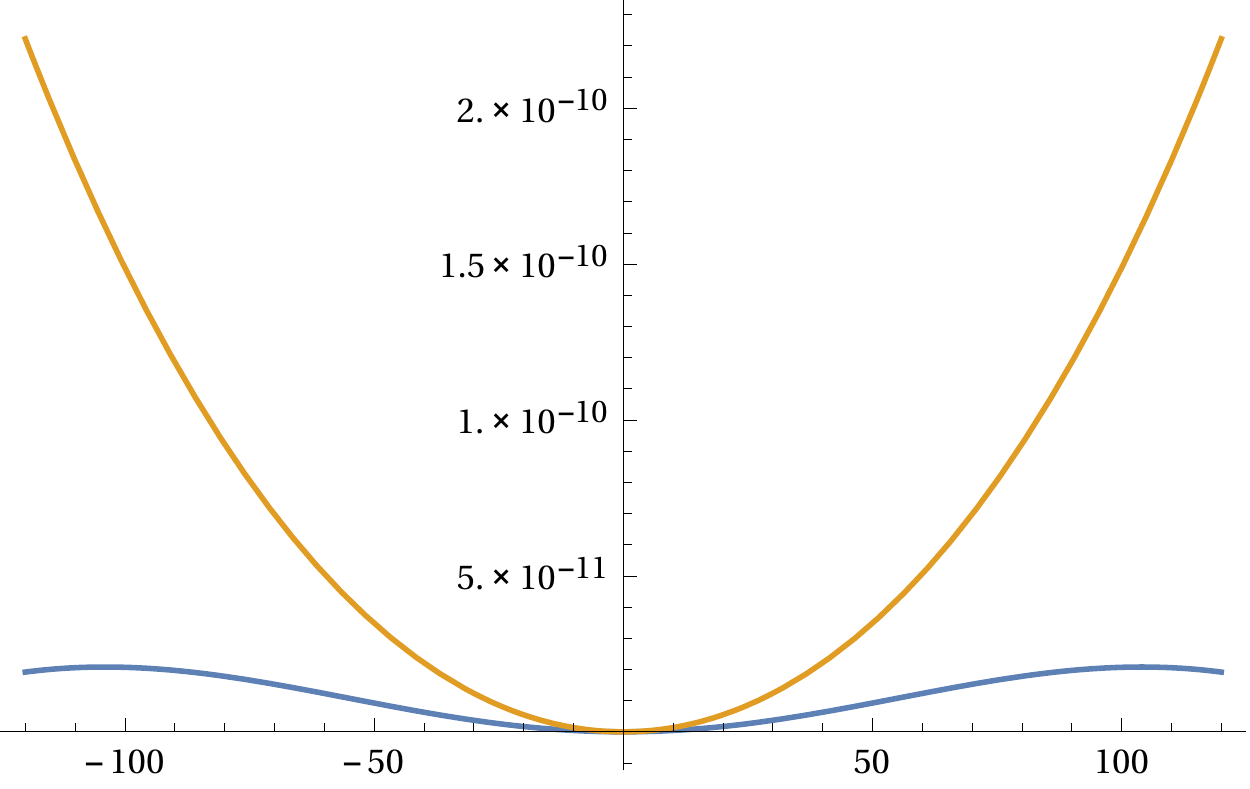}
 \begin{picture}(0,0)
  \put(-113,140){\footnotesize $V$}
   \put(0,10){\footnotesize $\theta$}
  \end{picture}
\caption{\label{fig:Vback2} Backreacted (blue line) and quadratic potential in units $\frac{M_{\rm Pl}^4}{4\pi}$ 
given by eq.~\eqref{vback2}
for $h = 1/220$,  $\tilde{\mff} = 1/1810$, $\mff = 6/49$, $q = 1/8$, $g = 1/10$, $p = 1/10000$ and $\lambda = 5$.}
\end{figure}
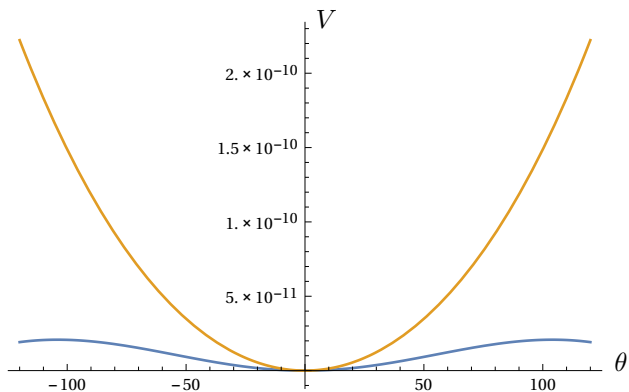 

For this limit situation we have, as shown in figure \ref{fig:Mratio}, a greater separation between
the KK scale and the string scale, while the vev's for the moduli are kept in the perturbative regime. 
The effective scalar potential for $\lambda = 5$ has the form  \eqref{vback2} with coefficients
$B \cdot 10^{14} = \{ 1.3607, 1.2675 \cdot 10^{-5} \}$, so that it
effectively behaves as a 
quadratic potential near the origin (see figure \ref{fig:Vback2}).
In this case a Minkowski vacuum is obtained taking $\delta \cdot 10^{7} = 4.2004$. 

As expected, for lower values of $\lambda$ the flattening effect of the backreaction becomes more important. 
In table \ref{tab:param2}
we display the relevant cosmological parameters for $\lambda = 5$. 
We find a  similar pattern as in the model presented in section \ref{ss:numerical}, but
now the number of e-foldings  is fairly large, while the
tensor-to-scalar ratio is almost as low as for the Starobinsky model.
That by decreasing $\lambda$ the model changes from
quadratic to plateau-like inflation has also been observed in \cite{Blumenhagen:2015qda}.

\begin {table}[htbp]
\begin{center}
\begin{tabular}{ |c | c |c |c |  }
  \hline
 Parameter & Value \\
 \hline
 $\Delta c$ 		&  $86$ $M_{\rm Pl}$ \\
 $N_e$ 			& $125$		\\
 $r$			& $0.007$	 		\\
 $n_s$			& $0.9667$		\\
 $\mathcal{P}$		& $2.14 \cdot 10^{-9} \,$ 	\\
 $M_{\rm s}$		&  $1.37 \cdot 10^{17} \,$GeV  \\
 $M_{\rm KK}$		& $1.76\cdot 10^{16} \,$GeV  \\
 $M_{\rm inf}$		& $ 2.74 \cdot 10^{15}\, $GeV   \\
 $M_{\rm mod}$		& $\{7.91,3.11,1.65;6.68,2.12\} \cdot  10^{14}\, $GeV   \\
 $H_{\rm inf}$		& $2.08 \cdot 10^{13} \,$GeV \\
 $M_{\theta}$		&  $ 4.69 \cdot 10^{12}\, $GeV\\
\hline
\end{tabular} 
\caption{\label{tab:param2} Summary of inflationary parameters for $\lambda = 5$.}
\end{center}
\end{table}

\newpage


\clearpage
\bibliography{references}  
\bibliographystyle{utphys}


\end{document}